\documentclass[twocolumn,showpacs,preprintnumbers,amsmath,amssymb]{revtex4}

\usepackage{graphicx}
\usepackage{dcolumn}
\usepackage{bm}

\begin{document}

\title{Analytical solution of the Thomas-Fermi equation for atoms}
\author{M. Oulne}
\email{oulne@ucam.ac.ma} \affiliation{ Laboratoire de Physique des
Hautes Energies et d'Astrophysique, Facult\'e des Sciences
Semlalia, Universit\'e Cadi Ayyad Marrakech, BP : 2390, Morocco.}

\date{\today}

\begin{abstract}
An approximate analytical solution of the Thomas-Fermi equation
for neutral atoms is obtained, using the Ritz variational method,
which reproduces accurately the numerical solution, in the range
$0\leq x\leq50$, and its derivative at $x=0$. The proposed
solution is used to calculate the total ionization energies of
heavy atoms. The obtained results are in good agreement with the
Hartree-Fock ones and better than those obtained from previously
proposed trial functions by other authors.
\end{abstract}

\pacs{ 31.15.Bs}
\maketitle

\section{INTRODUCTION} 
Since the first works of Thomas and Fermi \cite{1}, there have
been many attempts to construct an approximate analytical solution
of the Thomas-Fermi equation for atoms \cite{1}. E.Roberts
\cite{2} suggested a one-parameter trial function:
\begin{eqnarray}
  \phi_{1}(x)=(1+\eta\sqrt{x})e^{-\eta\sqrt{x}},
\end{eqnarray}
where $\eta=1.905$ and Csavinsky \cite{3} has proposed a
two-parameters trial function:
\begin{eqnarray}
  \phi_{2}(x)=(a_{0}e^{-\alpha_{0}x}+b_{0}e^{-\beta_{0}x})^2,
\end{eqnarray}
where $a_{0}=0.7218337$, $\alpha_{0}=0.1782559$, $b_{0}=0.2781663$
and $\beta_{0}=1.759339$. Later,  Kesarwani and Varshni \cite{4}
have suggested:
\begin{eqnarray}
\phi_{3}=(ae^{-\alpha x}+be^{-\beta x}+ce^{-\gamma x})^2,
\end{eqnarray}
where $a=0.52495$, $\alpha=0.12062$, $b=0.43505$, $\beta=0.84795$,
$c=0.04$ and $\gamma=6.7469$. The equations (2) and (3) are
obtained by making use of an equivalent Firsov's variational
principle \cite{5}. The equation (1) has been modified by Wu \cite
{6} in the following form:
\begin{eqnarray}
\phi_{4}=(1+m\sqrt{x}+nx)^2e^{-m\sqrt{x}},
\end{eqnarray}
where $m=1.14837$ and $n=4.0187.10^{-6}$. Recently,  M. Desaix et
al.\cite{7} have proposed the following expression:
\begin{eqnarray}
\phi_{5}=\frac{1}{(1+(kx)^\beta)^\alpha},
\end{eqnarray}
where $k=0.4835$, $\alpha=2.098$ and $\beta=0.9238$. Moreover,
other attempts have been conducted to solve this problem
\cite{8,10}. But, all of these proposed trial functions cannot
reproduce well the numerical solution of the Thomas-Fermi equation
\cite{11} and its derivative at $x=0$. They didn't prove efficient
when used to calculate the total ionization energy of heavy atoms.
In the present work, we propose a new trial function, constructed
on the basis of the Wu \cite{6} function, which reproduces
correctly the numerical solution of the Thomas-Fermi equation
\cite{11}. It also gives more precise results for the total
ionization energies of heavy atoms in comparison with the
previously proposed approximate solutions.

\section{THEORY}
The Thomas-Fermi method consists in considering that all
electrons of an atom are subject to the same conditions: each
electron, subject to the energy conservation law, has a potential
energy $e\Phi(r)$ where $\Phi(r)$ is the mean value of the
potential owed to the nucleus and all other electrons. The
electronic charge density $\rho(r)$ and the potential $\Phi(r)$
are related via the Poisson equation:
\begin{eqnarray}
\frac{1}{r}\frac{d^{2}}{dr^{2}}(r\Phi(r))=-4\pi\rho(r),
\end{eqnarray}
assuming that $\rho(r)$ and $\Phi(r)$ are spherically symmetric.
The energy conservation law applied to an electron in the atom
gives the following relation:
\begin{equation}
\frac{p^{2}}{2m}-e\Phi(r)=E,
\end{equation}
From the equation (7), we can obtain the maximum of the electron
impulsion:
\begin{eqnarray}
p=\sqrt{2me\Phi(r)},
\end{eqnarray}
where $\Phi(r)$ has to satisfy the boundary conditions:
\begin{eqnarray}
\Phi(R)=0,\left(\frac{d\Phi(r)}{dr}\right)
_{R}=\left[\frac{d}{dr}(\frac{eZ}{r})\right]_{R}=-\frac{eZ}{R^{2}},
\end{eqnarray}
where R is the radius of a sphere representing the atom. By
considering that the contribution of the electrons situated near
the nucleus to the potential $\Phi(r)$ is null, we obtain another
boundary condition:
\begin{eqnarray}
r\Phi(r)\rightarrow eZ \;for\; r\rightarrow 0,
\end{eqnarray}
The electronic charge density is defined by the relation:
\begin{eqnarray}
\rho=-\frac{8\pi e}{3}\left(\frac{p}{h}\right)^{2},
\end{eqnarray}
where p is the electron impulsion and h the Planck's constant. By
combining the relations (8) and (11), we obtain the following
expression for the charge density:
\begin{eqnarray}
\rho=-\frac{8\pi}{3}\frac{e}{h^{3}}\lbrack2me\Phi(r)\rbrack^{3/2},
\end{eqnarray}
To get rid of the numerical constants in the
equations, one can perform the following changes:
\begin{eqnarray}
x=\frac{r}{a},\phi(x)=\frac{1}{Ze}r\Phi(r),
\end{eqnarray}
with $a=a_{B}(\frac{9\pi^{2}}{128Z})^{1/3}$, where
$a_{B}=\frac{h^{2}}{4\pi^{2}me^{2}}$ is the first Bohr radius of
the hydrogen atom and r is the distance from the nucleus. With
these changes, we get from the equations (6) and (13) the
differential equation of Thomas-Fermi \cite{1}:
\begin{eqnarray}
\frac{d^{2}\phi}{dx^{2}}=\sqrt{\frac{\phi^{3}}{x}},
\end{eqnarray}
with the boundary and subsidiary conditions, obtained from the
equations (9) and (10):
\begin{eqnarray}
\phi(0)=1,\phi(\infty) = 0,\left(
\frac{d\phi}{dx}\right)_{x\rightarrow\infty}=0,
\end{eqnarray}
In this case, the charge density becomes:
\begin{eqnarray}
\rho=\frac{Z}{4\pi a^{3}}\left(\frac{\phi}{x}\right)^{3/2},
\end{eqnarray}
and  must satisfy the condition on the particles number:
\begin{eqnarray}
\int\rho dv=Z,
\end{eqnarray}
where Z is the number of electrons in neutral atom and dv is the
volume element. The use of the variational principle to the
lagrangian:
\begin{equation}
L(\phi)=\int_{0}^{\infty} Fdx,
\end{equation}
where:
\begin{equation}
F(\phi,\phi^{'},x)=\frac{1}{2}\left(\frac{d\phi}{dx}\right)^{2}+\frac{2}{5}
\left(\frac{\phi^{5/2}}{\sqrt{x}}\right),
\end{equation}
is equivalent to the equation (14) since substitution of the
functional (19) into the Euler-Lagrange equation:
\begin{eqnarray}
\frac{d}{dx}\left(\frac{\partial
F}{\partial\phi^{'}}\right)-\frac{\partial F}{\partial\phi}=0,
\end{eqnarray}
leads to the Thomas-Fermi equation (14). While solving the
Thomas-Fermi problem by using the variational principle, we can
assume an infinite number of trial functions which depend on
different variational parameters. In this paper, we propose a
trial function which depends on three parameters $\alpha$,
$\beta$ and $\gamma$:
\begin{equation}
  \phi(x)=(1+\alpha\sqrt{x}+\beta
  xe^{-\gamma\sqrt{x}})^{2}e^{-2\alpha\sqrt{x}},
\end{equation}
\begin{figure}
\includegraphics[height=10cm,width=8cm]{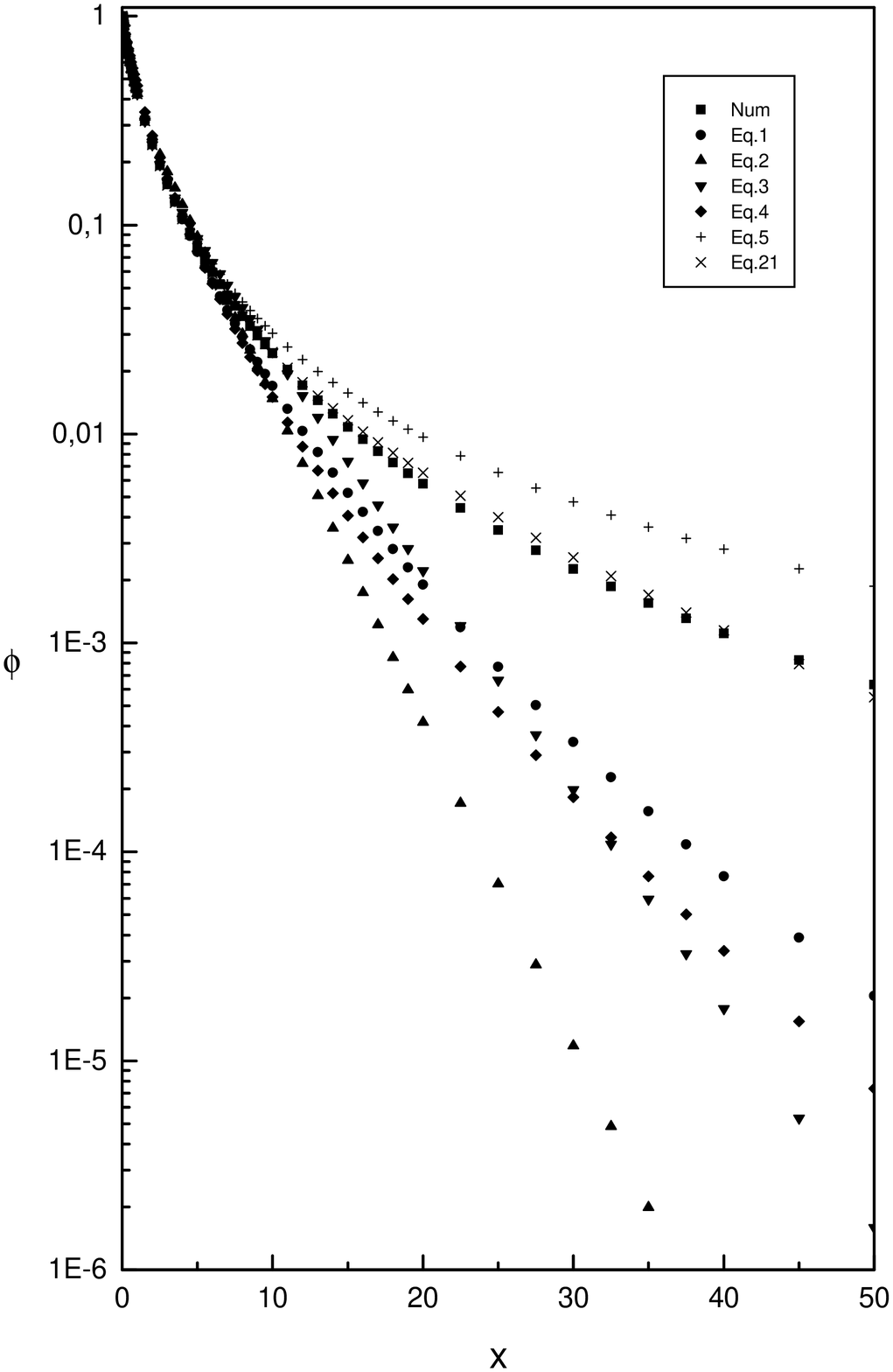}
\caption{\label{fig:epsart} Comparison of $\phi$ from Eqs.(1),
(2), (3), (4), (5) and (21).}
\end{figure}
\- After inserting the equation (21) into the equations (19) and
(18), the lagrangian $L(\phi)$ transforms into an algebraic
function $L(\alpha, \beta,\gamma)$ of the variational parameters
$\alpha$, $\beta$ and $\gamma$ and the Thomas-Fermi problem turns
into minimizing $L(\alpha, \beta,\gamma)$ with respect to these
parameters subject to the constraint (17) which is taken into
account through a Lagrange multiplier. All calculations, in this
work, are performed with the software Maple Release 9.
\section{RESULTS}
The optimum values of the variational parameters $\alpha$, $\beta$
and $\gamma$, obtained by minimizing the lagrangian (18) taking
into account the subsidiary condition (17), are respectively equal
to 0.7280642371, -0.5430794693 and 0.3612163121. The obtained
trial function (Eq.(21)), with these universal parameters,
reproduces accurately the numerical solution \cite{11} of the
Thomas-Fermi equation (14), in the range $0\leq x\leq50$, in
comparison with the equations (1), (2), (3), (4) and (5) as it is
shown in Fig. 1 and Tab. I. The mean error of our calculations,
calculated on 67 points in the range $0\leq x\leq50$ with respect
to the numerical solution, is about 2 \% , while the other
calculations have a mean error greater than 17 \%.

\begin{figure}
\includegraphics[height=10cm,width=8cm]{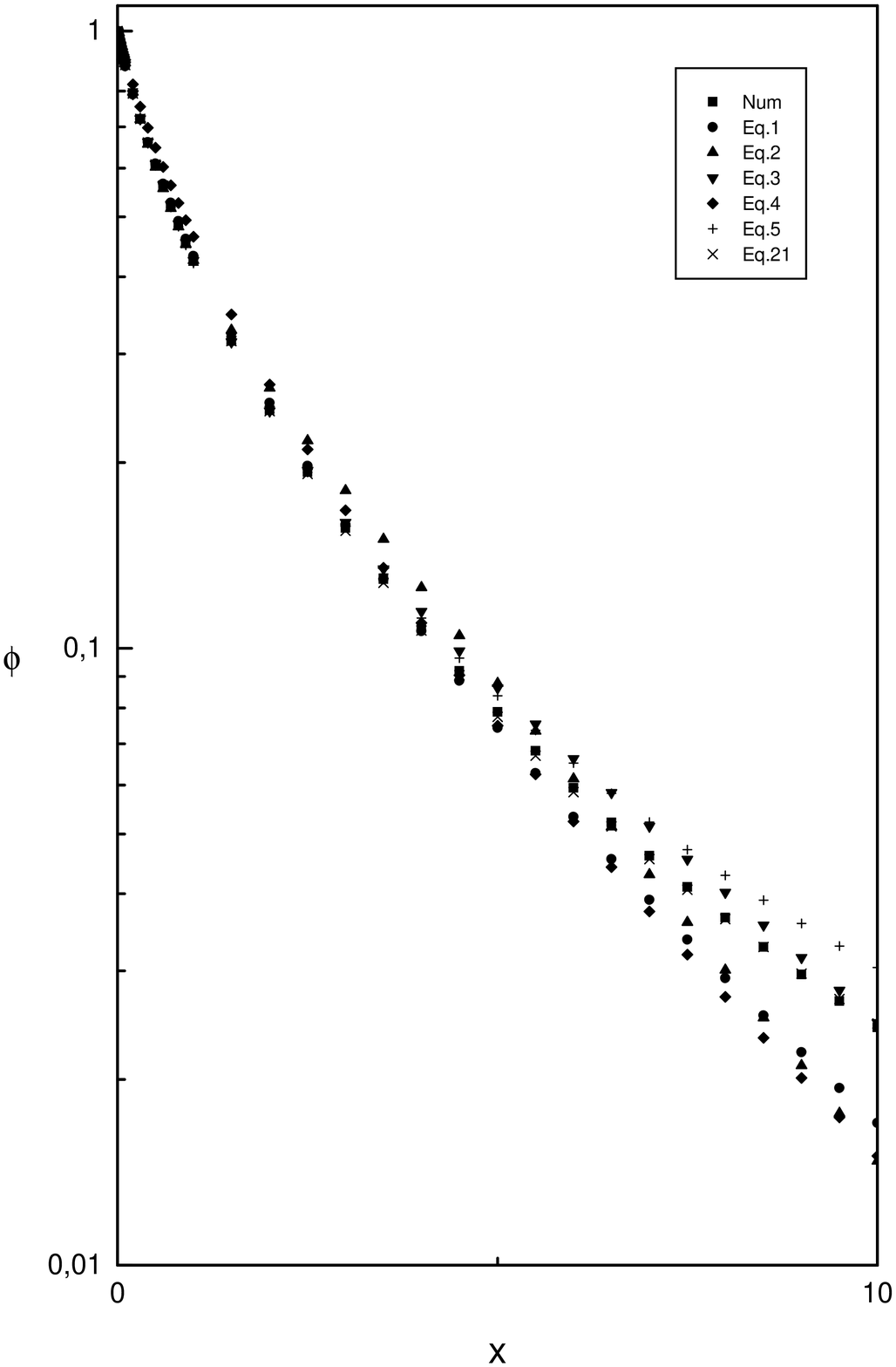}
\caption{\label{fig:epsart} Comparison of $\phi$ from Eqs.(1),
(2), (3), (4), (5) and (21) in the main region of the screening
potential.}
\end{figure}
\begin{table*}
\caption{\label{tab:table1}Comparison of the values of $\phi$ from
numerical solution and equations (1), (2), (3), (4), (5) and
(21). R.E(\%) is the relative error with respect to the numerical
solution}
\begin{ruledtabular}
\begin{tabular}{ccccccccccccccc}
x&\multicolumn{1}{c}{Num}&\multicolumn{1}{c}{Eq.(1)}&\multicolumn{1}{c}{R.E(\%)}&
\multicolumn{1}{c}{Eq.(2)}&\multicolumn{1}{c}{R.E(\%)}&\multicolumn{1}{c}{Eq.(3)}&
\multicolumn{1}{c}{R.E(\%)}&\multicolumn{1}{c}{Eq.(4)}&\multicolumn{1}{c}{R.E(\%)}&
\multicolumn{1}{c}{Eq.(5)}&\multicolumn{1}{c}{R.E(\%)}&\multicolumn{1}{c}{Eq.(21)}&
\multicolumn{1}{c}{R.E(\%)}\\
\hline

0   &   1   &   1   &   0.00    &   1   &   0.00    &   1   &   0.00    &   1   &   0.00    &   1   &   0.00    &   1   &   0.00    \\
0.001   &   0.9985  &   0.9983  &   -0.02   &   0.9988  &   0.03    &   0.9986  &   0.01    &   0.9987  &   0.02    &   0.9982  &   -0.03   &   0.9984  &   -0.01   \\
0.002   &   0.9969  &   0.9966  &   -0.03   &   0.9975  &   0.06    &   0.9972  &   0.03    &   0.9975  &   0.06    &   0.9966  &   -0.03   &   0.9969  &   0.00    \\
0.003   &   0.9955  &   0.9949  &   -0.06   &   0.9963  &   0.08    &   0.9958  &   0.03    &   0.9962  &   0.07    &   0.9950  &   -0.05   &   0.9954  &   -0.01   \\
0.004   &   0.994   &   0.9933  &   -0.07   &   0.9951  &   0.11    &   0.9944  &   0.04    &   0.9950  &   0.10    &   0.9935  &   -0.05   &   0.9939  &   -0.01   \\
0.005   &   0.9925  &   0.9917  &   -0.08   &   0.9939  &   0.14    &   0.9930  &   0.05    &   0.9938  &   0.13    &   0.9920  &   -0.05   &   0.9924  &   -0.01   \\
0.006   &   0.9911  &   0.9901  &   -0.10   &   0.9926  &   0.15    &   0.9917  &   0.06    &   0.9926  &   0.15    &   0.9906  &   -0.05   &   0.9910  &   -0.01   \\
0.007   &   0.9897  &   0.9886  &   -0.11   &   0.9914  &   0.17    &   0.9903  &   0.06    &   0.9914  &   0.17    &   0.9891  &   -0.06   &   0.9895  &   -0.02   \\
0.008   &   0.9882  &   0.9870  &   -0.12   &   0.9902  &   0.20    &   0.9889  &   0.07    &   0.9902  &   0.20    &   0.9877  &   -0.05   &   0.9881  &   -0.01   \\
0.009   &   0.9868  &   0.9855  &   -0.13   &   0.9890  &   0.22    &   0.9876  &   0.08    &   0.9890  &   0.22    &   0.9863  &   -0.05   &   0.9867  &   -0.01   \\
0.01    &   0.9854  &   0.9840  &   -0.14   &   0.9878  &   0.24    &   0.9862  &   0.08    &   0.9878  &   0.25    &   0.9849  &   -0.05   &   0.9853  &   -0.01   \\
0.05    &   0.9352  &   0.9314  &   -0.41   &   0.9412  &   0.64    &   0.9357  &   0.06    &   0.9451  &   1.06    &   0.9359  &   0.07    &   0.9348  &   -0.05   \\

0.09    &   0.8919  &   0.8874  &   -0.51   &   0.8983  &   0.72    &   0.8914  &   -0.05   &   0.9076  &   1.76    &   0.8933  &   0.16    &   0.8913  &   -0.06   \\

0.4 &   0.6596  &   0.6609  &   0.19    &   0.6557  &   -0.59   &   0.6607  &   0.16    &   0.6972  &   5.71    &   0.6598  &   0.03    &   0.6601  &   0.08    \\

0.8 &   0.4849  &   0.4920  &   1.47    &   0.4816  &   -0.68   &   0.4867  &   0.36    &   0.5268  &   8.63    &   0.4821  &   -0.57   &   0.4858  &   0.19    \\
1.5 &   0.3148  &   0.3233  &   2.70    &   0.3276  &   4.06    &
0.3136  & -0.38 &   0.3476  &   10.43   &   0.3116  &   -1.01   &
0.3147  &   -0.03\\
5   &   0.0788  &   0.0743  &   -5.71   &   0.0877  &   11.25   &   0.0861  &   9.30    &   0.0749  &   -4.96   &   0.0838  &   6.34    &   0.0774  &   -1.73   \\
10  &   0.0243  &   0.0170  &   -30.06  &   0.0147  &   -39.33  &   0.0247  &   1.73    &   0.0150  &   -38.11  &   0.0304  &   25.01   &   0.0247  &   1.44    \\
15  &   0.0108  &   0.0052  &   -51.53  &   0.0025  &   -77.04  &
0.0074  &   -31.56& 0.0041& -62.31& 0.0157& 45.66&  0.0116& 7.78\\
20  &   0.00578 &   0.00190 &   -67.13  &   0.00042 &   -92.78  &   0.00221 &   -61.72  &   0.00130 &   -77.51  &   0.00965 &   66.93   &   0.00653 &   12.98   \\

37.5    &   0.00131 &   0.00011 &   -91.70  &   8.14E-07    &   -99.94  &   3.25E-05    &   -97.52  &   5.03E-05    &   -96.16  &   3.17E-03    &   141.61  &   0.00140 &   6.87    \\

45  &   8.28E-04    &   3.88E-05    &   -95.31  &   5.62E-08    &   -99.99  &   5.32E-06    &   -99.36  &   1.54E-05    &   -98.14  &   2.27E-03    &   174.18  &   7.92E-04    &   -4.37   \\
50  &   6.32E-04    &   2.04E-05    &   -96.77  &   9.45E-09    &   -100    &   1.59E-06    &   -99.75  &   7.36E-06    &   -98.84  &   1.87E-03    &   196.02  &   5.50E-04    &   -12.90  \\

\end{tabular}
\end{ruledtabular}
\end{table*}

In the main region of the screening potential of Thomas-Fermi
$(0\leq x\leq10)$, our function is even more precise than all
other proposed functions as one can see from Fig. 2 and Tab. I.
The mean error of our calculations, calculated on 47 points in
this region, is equal to 0.28 \%, while the Eq.(2) has a mean
error equal to 1.13 \% and the Eqs.(1), (3), (4) and (5) have a
mean error greater than 2.5 \%.

The derivative of our function (Eq.(21)) at $x = 0$ is equal to
-1.61623647 which is close to the numerical derivative:
-1.58807102 \cite{11}. The relative error is less than 2 \%, while
the equations (1), (2), (3) and (4) give a result with an error
greater than 11 \% with respect to the numerical derivative and
the
Eq.(5) has an infinite derivative at x = 0. \\
\- To test the efficiency of the different trial functions, given
by the equations (1), (2), (3), (4) and (21),  we have calculated
the total ionization energy of heavy atoms   following the
relation \cite{12}:
\begin{equation}
 E=\left(\frac{12}{7}\right)\left(\frac{2}{9\pi^{2}}\right)^{1/3}
 \left(\frac{d\phi}{dx}\right)_{x=0}Z^{7/3},
\end{equation}
in hartrees $(e^{2}/a_{B})$ and the obtained results, presented in
Tab. II, are compared with those of Hartree-Fock (HF) \cite{13}.
The Eq.(5) cannot be used because of its infinite derivative at
$x=0$. From Tab. II, one can see that our results are fairly
better than those obtained from the Eqs.(1), (2), (3) and (4). The
precision of our calculations rises with the atomic number Z, on
the contrary of the other calculations performed with the Eqs.(1),
(2), (3) and (4), so our trial function is more suited for heavy atoms.\\

\section{CONCLUSION}
The proposed new trial function (Eq.(21)) provides a more
satisfactory approximation for the solution of the Thomas-Fermi
equation for neutral atoms than all other previousely proposed
analytical solutions. The results obtained for the total
ionization energies of heavy atoms agree with the Hartree-Fock
data and are more precise than those calculated with the Eqs.(1),
(2), (3) and (4). The proposed solution (Eq.(21)) can be used to
calculate, with high precision, other atomic characteristics of
heavy atoms.
\begin{table*}
\caption{\label{tab:table2}Comparison of total ionization energies
in units $(e^{2}/a_{B})$ from HF and equations (1), (2), (3), (4)
and (21).}
\begin{ruledtabular}
\begin{tabular}{cccccccccccc}
Z&\multicolumn{1}{c}{HF}&\multicolumn{1}{c}{Eq.(1)}&\multicolumn{1}{c}{Errors(\%)}&
\multicolumn{1}{c}{Eq.(2)}&\multicolumn{1}{c}{Errors(\%)}&\multicolumn{1}{c}{Eq.(3)}&
\multicolumn{1}{c}{Errors(\%)}&\multicolumn{1}{c}{Eq.(4)}&\multicolumn{1}{c}{Errors(\%)}&
\multicolumn{1}{c}{Eq.(21)}&\multicolumn{1}{c}{Errors(\%)}\\\hline
92&28070&33562&19.6&22864&-18.5&25972&-7.5&24392&-13.1&29894&6.5\\
93&28866&34419&19.2&23448&-18.8&26636&-7.7&25015&-13.3&30658&6.2\\
94&29678&35289&18.9&24040&-19.0&27309&-8.0&25647&-13.6&31433&5.9\\
95&30506&36171&18.6&24641&-19.2&27992&-8.2&26288&-13.8&32219&5.6\\
96&31351&37066&18.2&25251&-19.5&28684&-8.5&26938&-14.1&33015&5.3 \\
97&32213&37973&17.9&25869&-19.7&29386&-8.8&27598&-14.3&33823&5.0 \\
98&33093&38893&17.5&26495&-19.9&30098&-9.1&28266&-14.6&34643&4.7 \\
99&33990&39825&17.2&27130&-20.2&30819&-9.3&28944&-14.8&35473&4.4 \\
100&34905&40770&16.8&27774&-20.4&31550&-9.6&29631&-15.1&36315&4.0 \\
101&35839&41727&16.4&28426&-20.7&32292&-9.9&30327&-15.4&37168&3.7 \\
102&36793&42698&16.0&29088&-20.9&33042&-10.2&31032&-15.7&38032&3.4 \\
103& 37766&43681&   15.7&    29757 &  -21.2&   33803&   -10.5&   31746&   -15.9 &  38908 &  3.0 \\
104& 38758 & 44677  & 15.3  &  30436  & -21.5&   34574  & -10.8  & 32470 &  -16.2 &  39795&   2.7 \\
105& 39772 &  45686 &  14.9 &   31123 &  -21.7 &  35355 &  -11.1  & 33203  & -16.5&   40694 &  2.3 \\
106 &40806 &  46707  & 14.5&    31819 &  -22.0  & 36145 &  -11.4 &  33946 &  -16.8 &  41604 &  2.0 \\
107& 41862 &  47742 &  14.0  &  32524&   -22.3  & 36946 &  -11.7 &  34698 &  -17.1&   42525 &  1.6 \\
108& 42941 &  48790  & 13.6 &   33238 &  -22.6 &  37757  & -12.1&   35459 &  -17.4 &  43458  & 1.2 \\
109& 44042   &49850   &13.2  &  33960 &  -22.9  & 38578 &  -12.4 &  36230  & -17.7  & 44403 &  0.8 \\

\end{tabular}
\end{ruledtabular}
\end{table*}

\end{document}